\documentclass[12pt]{article}
\usepackage{amsfonts}

\usepackage{amsmath,amssymb,epsf}

\newcommand{\R}{{\mathbb{R}}}
\newcommand{\Z}{{\mathbb{Z}}}

\newcommand{\be}{\begin{equation}}
\newcommand{\ba}{\begin{eqnarray}}
\newcommand{\ea}{\end{eqnarray}}
\newcommand{\ee}{\end{equation}}

\begin{document}

\begin{titlepage}
\thispagestyle{empty}
\begin{flushleft}
CITUSC/00-056 \hfill hep-th/0010101 \\
UTMS 2000-57\hfill October, 2000 \\
UT-912 \\
\end{flushleft}

\vskip 1.5 cm

\begin{center}
\noindent{\Large \textbf{Tachyon Condensation on Noncommutative Torus}}\\
\noindent{
 }\\
\renewcommand{\thefootnote}{\fnsymbol{footnote}}

\vskip 2cm
{\large \it
I. Bars$^{a,}$\footnote{e-mail address: bars@physics.usc.edu}, 
H. Kajiura$^{b,}$\footnote{e-mail address: kuzzy@ms.u-tokyo.ac.jp},  
Y. Matsuo$^{c,}$\footnote{e-mail address:
 matsuo@hep-th.phys.s.u-tokyo.ac.jp},
 T. Takayanagi$^{c,}$\footnote{e-mail address: takayana@hep-th.phys.s.u-tokyo.ac.jp} } \\
{\it
\noindent{ \bigskip }\\
$^a)$ Caltech-USC Center for Theoretical Physics and Department of Physics\\
University of Southern California, Los Angeles, CA 90089-2535 \\
\noindent{\smallskip}\\
$^b)$ Graduate School of Mathematical Sciences, University of Tokyo \\
Komaba 3-8-1, Meguro-ku, Tokyo 153-8914, Japan\\
\noindent{\smallskip  }\\
$^c)$ Department of Physics, Faculty of Science, University of Tokyo \\
Hongo 7-3-1, Bunkyo-ku, Tokyo 113-0034, Japan\\
\noindent{ \smallskip }\\
}
\bigskip
\end{center}
\begin{abstract}
We discuss noncommutative solitons on a noncommutative
torus and their application to
tachyon condensation. In the large $B$ limit, they can be exactly
described by the Powers-Rieffel projection operators known in the mathematical
literature. The resulting soliton spectrum is consistent with T-duality
and is surprisingly interesting. It is shown that an instability arises
for any D-branes, leading to the decay into many smaller D-branes.
This phenomenon is the consequence of the fact that $K$-homology
for type II von Neumann factor is labeled by $\R$.
\end{abstract}
\vfill
\end{titlepage}
\vfill
\setcounter{footnote}{0}
\renewcommand{\thefootnote}{\arabic{footnote}}
\newpage


\section{Introduction}

In recent developments of string theory, there are several key observations
which characterize the geometry of the string and D-branes.

One such idea is noncommutative (NC) geometry which arises very naturally
when the background $B_{\mu \nu }$ field is nonvanishing \cite{cds,dh,sw}.
One of the most intriguing aspects is that the $B$ field not only deforms
the classical commutative background, but also it sometimes smears the
singularity of the geometrical configuration and defines the smooth solution
which is not present in the commutative limit. One of the most interesting
examples is the $U(1)$ instanton solution \cite{ns,furu}. 
The existence of such a
solution is quite desirable since it describes the physical configuration of
D0-branes in a D4 world volume, which is expected in string theory.

$K$-theory gives another clue for understanding the geometry of D-branes
\cite{mm,witten1,horava}. The $K$-theory setting becomes necessary since the
massless modes on the D-brane not only describe the embedding in space-time
but also the vector bundle over the world volume. Combined with the idea of
tachyon condensation \cite{sen13,sen-1}, we need to take the formal
difference between two vector bundles. This is actually the essence of the
topological $K^{0}$ group. Witten claimed that all the BPS D-brane charges
of type IIB string theory can be labeled by the $K^{0}$ group of target
space \cite{witten1}. A similar idea for type IIA was developed by Horava,
and in this case the classification is given by $K^{1}$ \cite{horava}.

A natural generalization of the $K$-groups to NC geometry is given by the $K$
homologies of the $C^{\ast }$-algebras. In the passage from commutative to
noncommutative description, the ring of functions on some topological space
is replaced by an abstract NC algebra $\mathcal{A}$. The idea of vector
bundles is generalized to projection operators of $\mathcal{A}\otimes M_{n}({%
\Bbb{C}})$. Physically, the matrix algebra $M_{n}({\Bbb{C}})$ describes
several D-branes which share the same world volume. The formal difference
between the projectors gives the element of $K_{0}(\mathcal{A})$ \cite{wo,co}%
. On the other hand, $K_{1}(\mathcal{A})$ is described by $U(\mathcal{A})/U(%
\mathcal{A})_{0},$ where $U(\mathcal{A})$ is unitary matrices with entries
in $\mathcal{A}$, and $U(\mathcal{A})_{0}$ is their connected components
\cite{wo,co}. 

After the discovery of the NC soliton \cite{gop}, such an abstract machinery
of $K$-homologies (especially $K_{0}(\mathcal{A})$) becomes relevant to
describe D-brane physics \cite{harvey}-\cite{rey}. This is because the NC
soliton is given in terms of projection operators. When the idea was applied
to tachyon condensation, the ``geometry'' of the projection operator \cite
{matsuo,hm} turns out to give that of D-branes.

A natural question is whether such noncommutative description gives novel
physics which does not appear in the commutative limit. To develop some such
ideas, it is natural to examine the $C^{\ast }$-algebra which has a richer
structure than the algebra of harmonic oscillators. In this paper, we
analyze the quantum torus \cite{co} as an example and find some new physical
phenomena. This arises from the following two facts:

\begin{enumerate}
\item  The NC torus describes a compact space as compared to the usual
infinitely extended Moyal plane. In NC geometry, the point like objects such
as D0-branes are forced to have a finite size. Since they are mutually
exclusive in the NC description, we will have a bound on the maximum
population which can live on the finite world.

\item  Unlike the algebra of harmonic oscillators, the algebra of the NC
torus is categorized as type II$_{1}$ von Neumann factor \cite{co}. This
means that the $K$-homology is not labeled by ${\Bbb{Z}}$ but by ${\Bbb{R}}$
and it is very interesting to ask how such charges may be interpreted, and
what is the physical consequence. This is in striking contrast to the
algebra of type I factor \cite{co} or matrix algebra.
\end{enumerate}

We find that these two issues are deeply interwoven and lead to physical
consequences: When the D0-branes are accumulated to some bound, they
can not be the consistent solutions but are reshaped
into smaller quanta which may be interpreted as bound states of
D0-branes and D2-branes. The existence of such bound states lead to a much
more complex spectrum. Indeed the mass spectrum is discrete but dense in $%
[0,1],$ which leads to an instability of the system. We will argue that such
behavior is more generally shared by D-branes which live on various compact
spaces.


%

\section{Review of noncommutative soliton}

Let us start with a brief review of the basic idea of the NC soliton on the
NC plane \cite{gop}. Consider the scalar field theory living in $2+1$
dimensions with noncommutativity in the two spatial directions $x^{1},x^{2}$%
. The energy of the system is given by,
\begin{equation}
E=\int_{R^{2}}d^{2}x\left( (\partial _{\vec{x}}\phi )^{2}+\Theta V(\ast \phi
)\right) \,\,,
\end{equation}
where $\Theta $ describes the noncommutativity in the Moyal plane. The $\ast
$ product is normalized to
\[
A(x)\ast B(x)=e^{\frac{i}{2}\epsilon _{ij}\partial _{\xi ^{i}}\partial
_{\eta ^{j}}}A(x+\xi )B(x+\eta )|_{\xi =\eta =0}\,\,.
\]
By using the Weyl correspondence, the functions on the Moyal plane can be
mapped to the linear operators acting on the Hilbert space $\mathcal{H}$ of
the harmonic oscillators, which we denote as $\mathcal{A}=\mathcal{B}(%
\mathcal{H})$. The integration with respect to the NC coordinates $\vec{x}$
is translated into the trace in $\mathcal{A}$.

In the large $\Theta $ limit the kinetic term can be neglected and the
stable configuration can be achieved by minimizing the potential part. The
main observation of \cite{gop} was that the projection operators in $%
\mathcal{A}$ gives the soliton states. Namely given the mutually orthogonal
projections $\phi _{i}\in \mathcal{A}$ (namely $\phi _{i}\cdot \phi
_{j}=\delta _{ij}\phi _{i}$) and a set of the critical values $\left\{
\lambda _{i}\right\} $ which solve $\frac{\partial V(\lambda )}{\partial
\lambda }=0$, one may construct the soliton solution as,
\begin{equation}
\phi (x)=\sum_{i}\lambda _{i}\phi _{i}(x).
\end{equation}
In particular, the level $k$ solution is given by projection operators $\phi
_{k}$ up to unitary transformations $\Lambda \in \mathcal{B}(\mathcal{H})$,
as follows
\begin{equation}
\phi _{k}(x)=\Lambda ^{\dagger }\left[ \sum_{\ell =0}^{k-1}\frac{1}{\ell !}%
(a^{\dagger })^{\ell }|0\rangle \langle 0|(a)^{\ell }\right] \Lambda ,
\end{equation}
where we have defined $a=\frac{1}{\sqrt{2}}(x_{1}-ix_{2})$. For such
solution, say $\phi =\lambda \phi _{k}$, the energy is given as,
\begin{equation}
E=\theta V(\lambda )\,k\,\,.  \label{energy}
\end{equation}

This idea was later applied to string theory \cite{harvey,das}, and the
scalar field was identified with the tachyon field. The NC soliton was then
interpreted as the D-branes which appear after tachyon condensation. In this
context, the integer $k$ was identified as the number of D-branes, and
formula (\ref{energy}) was interpreted as giving the correct D-brane
tension. Furthermore, in \cite{harvey} the gauge symmetry on such D-branes
was shown to be $U(k)$. In this interpretation, it was essential that the
level $k$ projector could be decomposed into $k$ mutually orthogonal
projectors, and each projector was identified with one D-brane. The open
string wave function $\Psi $ can then be projected into pieces $\phi
_{i}\Psi \phi _{j}$ which represent the open strings that interpolate the $i$%
-th and $j$-th branes. Similar results can also be obtained in the case of
the NC solitons on a fuzzy sphere \cite{hnt}.


\section{Noncommutative torus}


In this paper, we extend this framework by replacing the Moyal plane by the
NC torus. This can be most easily achieved by replacing the algebra $%
\mathcal{A}$ by $\mathcal{A}_{\theta }$ which is generated by two unitary
elements $U$ and $V$ satisfying the relation
\begin{equation}
UV=e^{-2\pi i\theta }VU\ .  \label{1}
\end{equation}
Geometrically $\mathcal{A}_{\theta }$ is related to the NC 2-torus ${{\Bbb{T}%
}}^{2}$ with radii $R_{1},R_{2}$ (rescaled to $1$ for convenience in the
following). Then $U,V$ are the exponentials of the noncommutative
coordinates, $\left( x_{1},x_{2}\right) \in $ ${{\Bbb{T}}}^{2},$ which
perform the translation around the two cycles of the torus
\begin{equation}
U=e^{ix_{1}},\ \ \ \ V=e^{ix_{2}}.
\end{equation}
With our normalization of coordinates, $\theta $ is a measure of the
magnetic flux through the torus$.$

When $\theta $ is rational, namely $\theta =q/p$ for mutually co-prime
integers $p,q$, the operators that correspond to making $p$ full
translations around either cycle of the torus $U^{p}$ or $V^{p}$ commute
with either $U$ or $V$. Then they act like the identity operator $%
U^{p}=V^{p}=\mathbf{1},$ which allows these generators to be expressed by
finite size $p\times p$ matrices,
\begin{equation}
U=\left(
\begin{array}{cccc}
1 & 0 & \cdots  & 0 \\
0 & \omega  & \cdots  & 0 \\
\vdots  & \vdots  & \ddots  & \vdots  \\
0 & 0 & \cdots  & \omega ^{p-1}
\end{array}
\right) \,,\quad V=\left(
\begin{array}{cccc}
0 & 1 & \cdots  & 0 \\
\vdots  & \vdots  & \ddots  & \vdots  \\
0 & 0 & \cdots  & 1 \\
1 & 0 & \cdots  & 0
\end{array}
\right) \,,\quad \omega =e^{2\pi i\theta }\,.
\end{equation}
Actually each entry in these matrices could be considered as blocks
multiplied by the identity operator. On the other hand, when $\theta $ is
irrational, there is no representation by finite size matrices but they
should be expressed in the infinite dimensional Hilbert space \cite{pv-2,lls}%
. A sequence of finite matrices associated with rational $\theta ,$ that
approach infinite matrices as $\theta $ approaches an irrational number is
described in \cite{lls}.%

{}These two situations appear to be very different in terms of matrix
representations, although in principle the physics should not be sensitive
to small variations from rational to irrational values of $\theta $.
Mathematically, whereas the irrational case describes the quantum torus, the
geometry for the rational case naively appears as if it is collapsed to a
finite number of points. 

In the rational case, a lattice version of the NC torus can also be
formulated \cite{bars} which helps to define a field theory (with a cutoff)
on a latticized NC torus with $N\times N$ cells. Consider $N^{2}$ discrete
points $\left( x^{1},x^{2}\right) $ on the torus that are labelled by $%
\left( j_{1},j_{2}\right) $ with $j_{1,}j_{2}=1,\cdots ,N,$ while the
positions (eigenvalues of the non-commutative operators) are defined by $%
x_{1,2}=aj_{1,2}$ where $a$ is the lattice distance. Discrete translations
connect these points to each other. The smallest translations $u,v$ in the
two directions correspond to the $N$-th root of the translations above $%
u=U^{1/N}$ and $v=V^{1/N},$ so these enter as the basic elements of the
algebra on the discretized torus. Since these are non-commuting operators
one may work in a basis in which one of them is diagonal, $u|x^{1}>=\tilde{%
\omega}^{j_{1}}|x^{1}>$ where $\tilde{\omega}$ is the $N$'th root of the
phase above $\tilde{\omega}=\omega ^{1/N}=e^{2\pi i\theta /N}.$  In this
basis $u$ is a diagonal matrix (clock) while $v$ is a periodic shift matrix.
Since $p$ full translations around the two cycles of the torus correspond to
$u^{Np}=v^{Np}=1,$ the matrix notation of $u,v$ is given in terms of $\left(
Np\right) \times \left( Np\right) $ matrices of the type above, instead of $%
p\times p$ matrices. In comparing to the lattice interpretation in \cite
{bars} we may use $n=Np$ and identify the $n\times n$ matrices $h,g,$ in
\cite{bars} as $u\rightarrow h$ and $v\rightarrow g$. These matrices replace
the $U$ and $V$ above when we discuss the latticized NC torus. Evidently $%
uv=vu\tilde{\omega},$ where $\tilde{\omega}=e^{i2\pi \tilde{\theta}}$ is
given by the magnetic flux $\tilde{\theta}=\theta /N$ through one plaquette
on the latticized torus.

A generic element $a\in \mathcal{A}_{\theta }$ can be expanded in the form
\begin{equation}
a=\sum_{m,n\in {\Z}}a_{mn}U^{m}V^{n}\ .
\end{equation}
For the rational case, the summation is limited to the range between $0$ and
$p-1$ (for the lattice version $U,V$ are replaced by $u,v$, and the
summation extends to the range between $0$ to $Np-1$). In the definition of
the NC soliton, the integration over NC space is replaced by the trace of
the $C^{\ast}$-algebra. For $\mathcal{A}_{\theta }$, one can define it as
\begin{equation}
\mbox{Tr}\,\,a:=a_{00},
\end{equation}
by using the above expansion. This reduces to the conventional trace of the
matrix for the rational case (up to normalization). For the irrational
situation, one may confirm $\mbox{Tr}(ab)=\mbox{Tr}(ba)$ for any elements
which 
is a compact operator\footnote{%
That is, when $a_{mn}$ tend to zero faster than any powers of $\left|
n\right| +\left| m\right| $ as $\left| n\right| +\left| m\right| \rightarrow
\infty .$}.

\section{NC soliton on fuzzy torus and lattice}

In this geometrical background, we consider the D-brane systems for which
there is a tachyonic instability and investigate the solitonic configuration
of the tachyon field on them in the large $B$ limit.

First let us discuss a non-BPS D2-brane (unstable D2-brane) \cite{gab}
wrapping the NC two torus. We denote the real tachyon field on it as $T$ and
the tachyon potential as $V(T)$. The tachyon field can be expressed as a
power-series of the operator $U$ and $V$ (or $u,v$ in the lattice version)
\begin{equation}
T=\sum_{n,m\in {{\Z}}}T_{nm}U^{n}V^{m}.
\end{equation}
The integers $n,m$ are interpreted as discrete momenta so that $T_{nm}$ is
the tachyon field in momentum space. The tachyon field in position space $%
T\left( \vec{x}\right) $ is given via a finite Fourier transform of $T_{nm}$
\cite{bars}. In the large $B$ limit we can ignore the kinetic term of $T$
and its effective action involves only the potential term in the same way as
in \cite{harvey,das,witten2,hnt}. Thus we get the total energy $E$ as
follows
\begin{equation}
E(T)=\mbox{M}_{D2}\ \mbox{Tr}[V(T)],
\end{equation}
where $\mbox{M}_{D2}$ denotes the mass of the original D2-brane and the
trace is normalized as $Tr[{\mathbf{1}}]=1$.

We assume the potential function $V(t)\ \ (t\in {\Bbb{R}})$ reaches its
minimum value at $t_{\min }=0$ and its local maximum at $t=t_{\max }$.
According to Sen's scenario \cite{sen13,sen-1} we can set $V(0)=0$ and $%
V(t_{\max })=1$. The original D2-brane configuration corresponds to $%
T=t_{\max }\cdot {\mathbf{1}}$ and the complete tachyon condensation ($T$=0)
leads to the decay to the vacuum. The equation of motion for $T$ is
satisfied if
\begin{equation}
T^{2}=t_{\max }\ T.
\end{equation}
Thus we can identify the allowed tachyon field as the projection $P$ in the
algebra $\mathcal{A}_{\theta }$:
\begin{equation}
T=t_{\max }\ P.
\end{equation}
Below we will set $t_{\max }\ =1$ for simplicity.

At this point, we need precise knowledge of the projection operators in $%
\mathcal{A}_{\theta }$. For the rational case, in matrix notation, it is
quite trivial. We can define a rank $k$ projection $P_{k}$ as a matrix with $%
k$ entries of $1$ on the diagonal and zeroes for all other entries
\begin{equation}
P_{k}=diag(1,\cdots ,1,0,\cdots ,0)=\sum_{n=0}^{p-1}T_{n0}U^{n}\,,\quad
T_{n0}=\frac{1}{p}\frac{1-\omega ^{-nk}}{1-\omega ^{-n}}\,.
\end{equation}
In the lattice case we replace $U$ by $u$ and $\omega $ by $\tilde{\omega},$
while the summation extends to $Np-1,$ and the rank of the diagonal matrix
is $Np.$ Since the sum contains only the diagonal $U$'s (or $u$'s), the
tachyon field in momentum space $T_{n0}$ has zero momentum in the $x^{2}$
direction. Therefore, its Fourier transform to position space defines a
tachyon lump that has a strip-like configuration\footnote{%
A similar argument can also be given in the case of Moyal plane as discussed
in \cite{rey}.} unlike the point-like one in the GMS soliton. This is,
however, rather superficial. Clearly $P_{k}$ can be modified by conjugating
with a unitary operator $P_{k}\rightarrow \Lambda ^{\dagger }P_{k}\Lambda $
without changing idempotency $P_{k}^{2}=P_{k}$. After such a transformation,
the shape of the soliton solution in position or momentum space is different
from the original one. In the GMS case, the minimization of the kinetic term
for finite $\theta $ favors the point-like configuration \cite{gop}. We
expect that a similar argument can be applied here to select the point-like
configuration although we have not explicitly examined the corresponding $%
\Lambda $ in detail since it is not important for what follows.

In the above example, the rank $k$ is limited to the range $[0,p],$ leading
to the trace $\mbox{Tr}\,P_{k}=k/p\in \lbrack 0,1]$ (on the lattice the
range is $\left[ 0,Np\right] $ leading to $\mbox{Tr}\,P_{k}=k/Np\in \lbrack
0,1]$). Since $k$ is interpreted as the number of $D$-branes, there is a
limit on the number of $D$-branes that can fit on the torus. This is the
appearance of the finite size effect we mentioned at the beginning. Note
also that a similar effect can be seen in the case of the fuzzy sphere \cite
{hnt}, where the algebra is also equivalent to finite size matrices.

Unlike the noncompact situation, we already encountered an important
difference in the property of the NC soliton. However, the physics on NC
torus for irrational $\theta $ is more complicated and interesting as we
will see below.

\section{Powers-Rieffel projector}

The generalization to the irrational case is rather interesting. We may
still diagonalize $U$ in $x^{1}$ space 
$U|x^{1}\rangle=e^{2\pi ix^{1}}|x^{1}\rangle$ and
represent $V$ as a shift operator $V|x^{1}\rangle=|x^{1}+\theta \rangle$ but now let $%
\theta $ be irrational.

Most naively, one may construct projection operators in the form $P_{\kappa
}=f(U),$ by choosing the function $f(e^{2\pi ix^{1}})$ to be the periodic
step function that takes the values  $f=1$ for $0\leq x^{1}\leq \kappa $ and
$f=0$ for $\kappa \langle x^{1}\leq 1,$ within one period $x^{1}\in \left[ 0,1%
\right] ,$ for any $0\leq \kappa \leq 1.$ This satisfies $P_{\kappa
}^{2}=P_{\kappa }$ as a simple multiplication of the function
and self-adjointness $P_\kappa^\dagger=P_\kappa$. Calculating
the trace we find $\mbox{Tr}(P)=\int_{0}^{1}dx^{1}\langle x^{1}|P_{\kappa
}|x^{1}\rangle=\int_{0}^{1}dx^{1}f(e^{2\pi ix^{1}})=\kappa $. Unfortunately, this
contradicts the expected spectrum $K_{0}(\mathcal{A}_{\theta })={\Bbb{Z}}%
+\theta {\Bbb{Z}}$ since $\kappa $ is not quantized. Indeed such family of
solutions do not use the noncommutativity at all. They are not acceptable as
the noncommutative soliton since they are singular and unstable.

To get the regular solution, we need to use both $U$ and $V$, and
incorporate the noncommutativity. 
Such solutions can be constructed by slightly modifying the naive solution
we discussed above. As a first example consider
\begin{equation}
P_{\theta }=V^{\dagger }\left( {g}(U)\right) ^{\dagger }+f(U)+g(U)V.
\label{PR1}
\end{equation}
Notice that the $\theta $ on $P_{\theta }$ is now the noncommutativity
parameter rather than being arbitrary; we will later construct more general
projectors. Acting on position space $|x^{1}\rangle$, we require $\left( P_{\theta
}\right) ^{2}|x^{1}\rangle=P_{\theta }|x^{1}\rangle$.  This defines a projection in $%
\mathcal{A}_{\theta }$ if and only if $f$ and $g$ satisfy the following
relations
\begin{eqnarray}
&&g(e^{2\pi ix^{1}})g(e^{2\pi i(x^{1}+\theta )})=0\,\,,  \nonumber \\
&&g(e^{2\pi ix^{1}})[1-f(e^{2\pi ix^{1}})-f(e^{2\pi i(x^{1}+\theta
)})]=0\,\,,  \nonumber \\
&&f(e^{2\pi ix^{1}})[1-f(e^{2\pi ix^{1}})]=|g(e^{2\pi
ix^{1}})|^{2}+|g(e^{2\pi i(x^{1}-\theta )})|^{2}\ .  \label{PR}
\end{eqnarray}
An explicit form of $f,g$ which satisfy these relations are given as
follows. Choose any small $\epsilon >0$ such that $\epsilon <\theta $ and $%
\theta +\epsilon <1$, and let $F_\theta(x^{1})\equiv f(e^{2\pi ix^{1}})$ 
for one period be given in the range $x^{1}\in \left[ 0,1\right] $ by
\begin{equation}
F_{\theta }(x^{1})=\left\{
\begin{array}{lcl}
x^{1}/\epsilon  & \qquad  & x^{1}\in \lbrack 0,\epsilon ] \\
1 &  & x^{1}\in \lbrack \epsilon ,\theta ] \\
1-(x^{1}-\theta )/\epsilon  &  & x^{1}\in \lbrack \theta ,\theta +\epsilon ]
\\
0 &  & x^{1}\in \lbrack \theta +\epsilon ,1]
\end{array}
\right. \,\,,  \label{F}
\end{equation}
Then define $g$ for one period by
\begin{equation}
g(e^{2\pi ix^{1}})=
\begin{cases}
\sqrt{F_{\theta }(x^{1})(1-F_{\theta }(x^{1}))} & \ \ x^{1}\in \lbrack
0,\epsilon ], \\
0 & \ \ x^{1}\in \lbrack \epsilon ,1]\ .
\end{cases}
.
\end{equation}
The functions $f$ and $g,$ defined as the periodic extensions of the above,
satisfy the relation (\ref{PR}). This projection is called the
Powers-Rieffel (PR) projection \cite{ri-2}. It can be easily shown that
\begin{equation}
\mbox{Tr}\,P_{\theta }=\int_{0}^{1}dx^{1}\langle x^{1}|P_{\theta
}|x^{1}\rangle=\int_{0}^{1}dx^{1}F_{\theta }(x^{1})=\theta \,\,.
\end{equation}
The parameter $\epsilon $ plays the r\^{o}le of regularizing the solution.
In the $\epsilon \rightarrow 0$ limit, the PR projector approaches the naive
solution we considered at the beginning, but only for the quantized value of
$\kappa =\theta .$ Similarly, we will find solutions for all the expected
quantized values of the projector as we will see below.

In position space, this projection defines a strip on the torus, just as in
the matrix version in the previous section. If we normalize the size of the
torus to one, the area of the strip is $\theta $ when $\epsilon $ is very
small. This is the analogue of the rational situation where the soliton
occupies $1/p$ of the area of the torus (or $1/Np$ in the lattice version).

In general, it is known that the $K_{0}$-group $K_{0}(\mathcal{A}_{\theta })$
is labeled by ${{\Bbb{Z}}}+\theta {{\Bbb{Z}}}$ \cite{pv-2}. 
The projection operators associated with them should satisfy
\begin{equation}
\mbox{Tr}P_{n+m\theta }=n+m\theta ,\qquad (0\leq n+m\theta \leq 1).
\label{trace}
\end{equation}
Such general projections can be constructed by slightly modifying the
Powers-Rieffel projection (\ref{PR}). For example, to define the projection
for $m\theta <1$, we modify the generators $U$ and $V$ in (\ref{PR1}) by the
combination which will produce the NC parameter $m\theta $. The simplest
choices are replacing $(U,V)$ in (\ref{PR1}) by (i) $(U^{m},V)$ or (ii) $%
(U,V^{m})$. In both cases, they generate $\mathcal{A}_{m\theta }$ which is
embedded in $\mathcal{A}_{\theta }$. In the first choice, the PR projection
is described by a function with period $1/m$ with each lump spreading over
the same range $\theta $ (Figure 1a). On the other hand, in the second
choice, the period is invariant but the width of the lump is enlarged to $%
m\theta $ (Figure 1b). In either case, the total area occupied by the lump
is $m\theta $.

\begin{figure}[th]
\centerline{\epsfxsize=12cm \epsfbox{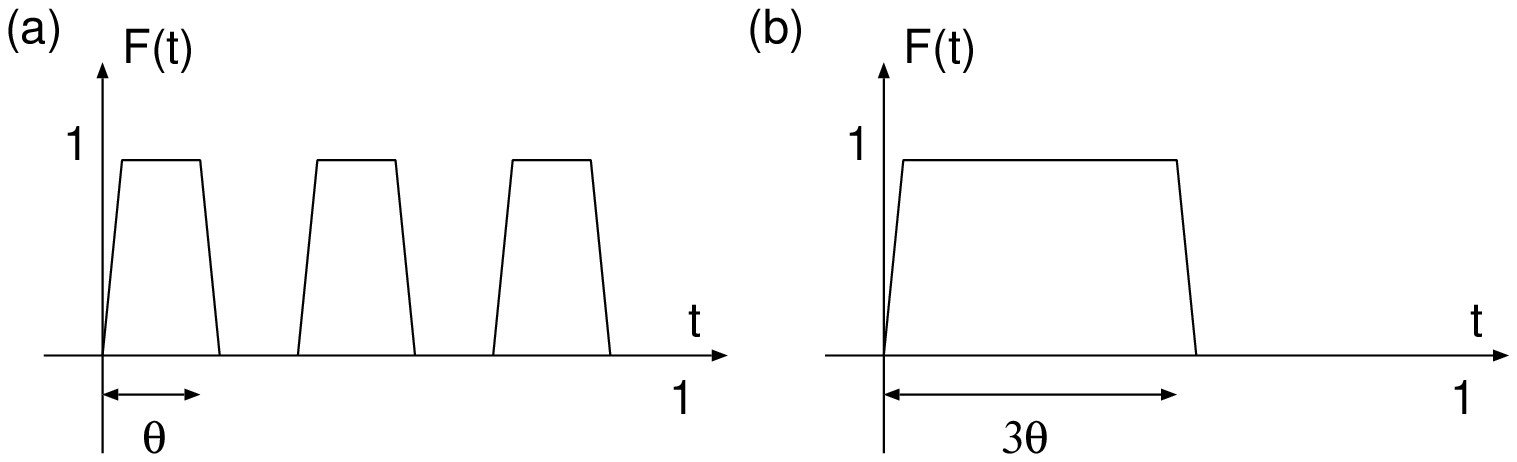}} \vskip 3mm
\caption{PR projection for $3\protect\theta $}
\end{figure}
In a sense, this is the analog of latticizing the torus by making $m$ steps,
and thus replacing the original $U$ by $u$ so that $u^{m}=U$, and then
renaming $u\rightarrow U$ (similarly for $V$ in case (ii)).

Such construction is valid as long as $m\theta <1$. If it exceeds 1, it
becomes inconsistent since $m\theta +\epsilon >1$. So we need to go back to
examine the condition (\ref{PR}) carefully. For simplicity we use the second
construction (ii). After a brief inspection, one notices that one may insert
$n+m\theta \in \lbrack 0,1]$ in the definition of $F\left( x^{1}\right) $
for one period in eq. (\ref{F}), but that $n$ drops out in eq.(\ref{PR}) due
to the periodicity of $f,g.$  Thus the general solution has the form
\begin{equation}\label{PR2}
P_{n+m\theta }=\left( V^{m}\right) ^{\dagger }\left( {g}(U)\right) ^{\dagger
}+f(U)+g(U)V^{m}
\end{equation}
with the information about $n$ inserted in the function $F_{n+m\theta }$ in (%
\ref{F}) with $\theta $ replaced by $0\leq n+m\theta \leq 1.$ The effect is
that the area of the NC soliton now shrinks to $n+m\theta $ for this new
solution. Again we see that for $\epsilon \rightarrow 0$ the solution is
similar to step function type solution $P_{\kappa }$ discussed at the
beginning of the section, but now $\kappa $ comes only in quantized values.

In a sense, this phenomenon may be physically understood as follows. One may
stuff D0-branes as much as possible while $m\theta <1$. After it reaches the
limit, the solution becomes inconsistent but may be transformed to the
smaller configuration which fits in the torus by subtracting the D2-brane
contribution. This family of solutions are interpreted as bound states of D2
and D0 branes in what follows.


\section{Spectrum and T-duality}

Up to now, we called the NC solitons D0-branes without specifying the
physical detail. {}From the trace formula (\ref{trace}), it is
straightforward to get the following mass spectrum
\begin{equation}
E(P_{n+m\theta })=(n+m\theta )\ \mbox{M}_{D2}.  \label{spectrum1}
\end{equation}
The most intriguing point is that this spectrum is dense in $0\leq E\leq %
\mbox{M}_{D2}$ for irrational $\theta $. What is the interpretation of these
excitations? A crucial observation is that $E=\theta \mbox{M}_{D2}$, which
is the mass for $(n,m)=(0,1)$, is the same as that of a D0-brane in the
large $B$ limit. Let us show this fact below. The mass of a non-BPS D2-brane
or a D0-brane is given by
\begin{eqnarray}
\mbox{M}_{D2} &=&\sqrt{2}\frac{R_{1}R_{2}}{g_{s}(\alpha ^{\prime })^{\frac{3%
}{2}}}\ \sqrt{1+(2\pi \alpha ^{\prime }B)^{2}}\ \sim \ \frac{2\sqrt{2}\pi
R_{1}R_{2}B}{g_{s}(\alpha ^{\prime })^{\frac{1}{2}}} \\
\mbox{M}_{D0} &=&\sqrt{2}\frac{1}{g_{s}(\alpha ^{\prime })^{\frac{1}{2}}},
\end{eqnarray}
where the factor $\sqrt{2}$ is peculiar to non-BPS D-branes. The
noncommutativity parameter $\theta $ on the torus is generated by the $B$%
-flux \cite{sw} and the relation between them is determined as
\begin{equation}
B=\frac{1}{2\pi R_{1}R_{2}\theta }.
\end{equation}
The large radius limit (Moyal plane) corresponds to the limit $\theta
\rightarrow 0$. The mass of a D0-brane can be written as $\mbox{M}%
_{D0}=\theta \mbox{M}_{D2}$ and thus we can represent the general spectrum (%
\ref{spectrum1}) as
\begin{equation}
E(P_{n+m\theta })=n\mbox{M}_{D2}+m\mbox{M}_{D0}\ \ \ (\ 0\leq n+m\theta \leq
1).  \label{spectrum2}
\end{equation}
The bound $0\leq n+m\theta \leq 1$ corresponds to the natural fact that
the mass of an object after tachyon condensation cannot exceed that of the
original D2-brane $T=P_{1}$.

Let us consider the interpretation of this spectrum. It is natural to
speculate that the object corresponding to $P_{n+m\theta }$ can be regarded
as a bound state of $n$ D2-branes and $m$ D0-branes. Indeed the mass of the
D2-D0 bound state is given by
\begin{eqnarray}
M_{(n,m)} &=&\sqrt{2}\frac{R_{1}R_{2}}{g_{s}(\alpha ^{\prime })^{\frac{3}{2}}%
}n\sqrt{1+(2\pi \alpha ^{\prime }B_{eff})^{2}},  \label{BImass} \\
B_{eff} &=&B+\frac{1}{2\pi R_{1}R_{2}}\frac{m}{n},
\end{eqnarray}
where $B_{eff}$ is defined as the effective $B$-field which includes the
flux due to $m$ D0-branes melting into the $n$ D2-branes. If we take $1\ll
(2\pi \alpha ^{\prime }B_{eff})^{2}$ then the mass formula (\ref{spectrum2})
is reproduced. Note that if $m<0$, we can interpret it as the annihilation
or tachyon condensation of $|m|$ (non-BPS) D0-branes; this interpretation is
natural if we remember the previous discussions of tachyon condensation. If $%
n<0$, then we cannot interpret the object from the viewpoint of the D2-brane
world-volume.

We argue that such objects really exist as we have already seen that they
appear very naturally in the study of tachyon condensation on NC torus. As
in the case of the BPS spectrum \cite{Ho}-\cite{K-S}, this is also confirmed
by examining T-duality as follows. The $SL(2,\Z)$ subgroup of
T-duality on the torus or Morita equivalence of $\mathcal{A}_{\theta }$ is
given as :
\begin{equation}
{\mathcal{A}}_{\theta }\sim {\mathcal{A}}_{\theta ^{\prime }}\leftrightarrow
\theta ^{\prime }=\frac{r+s\theta }{p+q\theta }\ \ \ \ (ps-qr=1,\ p,q,r,s\in
{{\Bbb{Z}}}).
\end{equation}
At the same time the open string coupling $G_{s}$ and the volume of torus $V$
have to be transformed \cite{BrMorZum,BrMor,sw} as
\begin{equation}
V^{\prime }=(p+q\theta )^{2}V,\ \ \ \ \ G_{s}^{\prime }=(p+q\theta )G_{s}.
\end{equation}
On the other hand, the mass of the object corresponding to $\mbox{P}%
_{n+m\theta }$ can be written in the open string language as
\begin{equation}\label{e-spectrum}
E({P}_{n+m\theta })=(n+m\theta )\frac{V}{G_{s}}.
\end{equation}
Then it is easy to see that if $(n,m)$ is transformed as
\begin{equation}
\left(
\begin{array}{c}
n^{\prime } \\
m^{\prime }\
\end{array}
\right) =\left(
\begin{array}{cc}
s & -r \\
-q & p
\end{array}
\right) \left(
\begin{array}{c}
n \\
m
\end{array}
\right) ,
\end{equation}
the energy spectrum (\ref{e-spectrum}) is invariant\footnote{
Precisely speaking, for some choice of $SL(2,\Z)$ transformation,
$n+m\theta$ can be negative and may not be interpreted as
the trace.  For such situation, $p+q\theta$ is also negative
and we should take the absolute value for them.
}.

Using this T-duality, we can always change the object into $k$ D0-branes,
where $k$ is the greatest common divisor of $(m,n)$. Then it is natural to
speculate that the object can be identified as $k$ bound states of D-branes
and there will be a $U(k)$ gauge theory on its world volume. However, this
bound state interpretation is a hasty conclusion and we will return to this
point later.

The above arguments apply to the D-branes in bosonic string theory almost in
the same way due to the factorization of the oscillator modes, and therefore
we omit any further discussion.

\section{Application to $D\bar D$-system}

Next we turn to the brane-antibrane system which consists of a D2-brane and
anti-D2 brane wrapped around the torus. In this case the tachyon field is a
complex scalar field $(T,\bar{T})$ and the gauge symmetry is $U(1)\times
U(1) $. The kinetic term of $(T,\bar{T})$ is again negligible in the large $%
B $ limit, and the total energy is given by the potential term as
\begin{eqnarray}
E(T,\bar{T}) &=&\mbox{M}_{D2}\mbox{Tr}\ [V(1-T\bar{T})+V(1-\bar{T}T)],
\nonumber \\
V(0) &=&0,\ \ \ V(1)=1,
\end{eqnarray}
where we have used the fact that the form of tachyon potential is
constrained due to the gauge symmetry, and that only the disk amplitude is
relevant in the leading order of $\frac{1}{g_{s}}>>1$. Note also $\mbox{M}%
_{D2}$ means the mass of a BPS D2-brane.

Let us define the operators $\Pi _{1},\Pi _{2}$ \cite{matsuo,hm} as
\[
\Pi _{1}=1-\bar{T}T,\ \ \ \Pi _{2}=1-T\bar{T}.
\]
The equation of motion is satisfied almost in the same way as the cases of
Moyal plane \cite{witten2,matsuo,hm} or fuzzy sphere \cite{hnt}, provided
the condition for the partial isometry
\begin{equation}
T\bar{T}T=T\,\,,\quad \bar{T}T\bar{T}=\bar{T}\,\,,
\end{equation}
holds and this shows that $\Pi _{1},\Pi _{2}$ are self-adjoint projections.

Therefore we get the following mass spectrum, which respects T-duality, and
again is dense,
\begin{eqnarray}
E &=&\mbox{M}_{D2}\mbox{Tr}(\Pi _{1}+\Pi _{2})  \nonumber \\
&=&\mbox{M}_{D2}[(n_{1}+n_{2})+(m_{1}+m_{2})\theta ],  \nonumber \\
&&(n_{i},m_{i}\in {{\Bbb{Z}}},0\leq n_{i}+m_{i}\theta \leq 1).  \label{mass}
\end{eqnarray}
This spectrum includes D0-branes and anti-D0 branes corresponding to $%
n_{i}=0 $ because the relation $\mbox{M}_{D0}=\theta \mbox{M}_{D2}$ still
holds. The mass formula is again consistent with that of BPS D2-D0 bound
states. In this case we can interpret ``$-|m|$ D0-branes'' as $m$ anti
D0-branes which are annihilated with some parts of $n$ D2-branes. We argue
the existence of the case $n<0$ is physical in the same way as in the
previous example of non-BPS D-branes. Although this mass formula does not
distinguish a D0 from an anti-D0, it is natural to conclude that the total
RR-charge of D0-brane is given by $m_{1}-m_{2}$\footnote{%
The topological nature of this charge is confirmed directly if one notes
that it is equivalent to the cyclic cocycle $\tau _{2}:K(\mathcal{A}_{\theta
})\rightarrow {\Z}$ \cite{co} as $\frac{1}{2\pi i}\tau
_{2}=m_{1}-m_{2} $.}. Note also that the index $\mbox{Index}(T)$, which has
been argued to be the D0-brane charge in NC $\R^2$ case \cite
{witten2,matsuo,hm}, is given in the case of our NC torus by
\begin{equation}
\mbox{Index}(T)=\mbox{Tr}(\Pi _{1}-\Pi
_{2})=(n_{1}-n_{2})+(m_{1}-m_{2})\theta .
\end{equation}
This shows that the index is not quantized, as is well-known for the NC
torus \cite{co}.

Finally let us discuss the relation between D-brane charges and $K$-groups.
As Witten argued in \cite{witten1} the D-brane charges in Type IIB theory
can be classified by the $K_{0}$ group, considering the tachyon condensation
in the brane-antibrane systems. In our case the suitable one is $K_{0}({%
\mathcal{A}}_{\theta })$ \cite{pv-2,ri-2,pv-1}:
\begin{equation}
K_{0}({\mathcal{A}}_{\theta })={\Z}+{\Z}\theta ,
\end{equation}
where the ordering is determined by the trace map (\ref{trace}). This fact
is easy to understand if one notes that an element of $K_{0}({\mathcal{A}}%
_{\theta })$ is a projection in $M_{n}\otimes {\mathcal{A}}_{\theta }$ and
one applies to that the Powers-Rieffel projection. Then we can conclude that
the tachyon field which is classified by $(n_{1},m_{1}),(n_{2},m_{2})$ in (%
\ref{mass}) corresponds to an element of the $K$-group as
\begin{equation}
(n_{1}-n_{2})+(m_{1}-m_{2})\theta \in K_{0}({\mathcal{A}}_{\theta }).
\end{equation}
One may understand the non-integrability of the D-brane charge
in the following fashion. As emphasized in \cite{mw}, 
the RR fields should take their values in the $K$-group.
In this sense, the $K$-group of NC torus is given by
$\Z+\theta\Z$ and we should take this value as the brane charge
as it is with the total ordering $n+m\theta \geq 0$. This is in
contrast with the commutative situation where $K$-group is described
as $\Z\oplus\Z$ where two $\Z$s can be interpreted as
D2 and D0 brane charges.  In our case, the distinction between
the two becomes obscure.  We also note that in the commutative
limit $\theta\rightarrow 0$, the group $\Z+\theta\Z$ reduces to
$\Z\oplus \Z$ as it should be.


\section{Instability}

Finally we discuss the instability of the branes. As we have argued, the
soliton associated with the PR projection for $n+m\theta $ should have the
gauge symmetry $U(k)$ with $k$ given by the greatest common divisor of $(n,m)
$. This statement is not so straightforward as it looks. It means that we
can define the open strings which interpolate the different D-branes. For
such open strings to exist, we need to have $k$ mutually orthogonal
projections $p_{i}$ which satisfy,
\begin{equation}
P_{n+m\theta }=\sum_{i=1}^{k}p_{i},\qquad p_{i}\cdot p_{j}=\delta
_{ij}p_{j}\,\,,
\end{equation}
as in the Moyal plane case. We can describe such decomposition by using the
Powers-Rieffel projection in the following way. Consider for simplicity the
case $P_{2\theta }$. We have already described that such a projection
operator can be constructed in two different ways. Let us consider the first
choice (i). Then $P_{2\theta }$ describes two lumps of the same shape.
Actually the first lump is identical to the PR projection operator for $%
\theta $. Suppose $\theta $ is small enough, $\theta <1/4$, so it can be
decomposed as two mutually orthogonal projections associated with two lumps.
Since the projection operator is split orthogonally, the gauge symmetry of
the D-brane which corresponds to $T=P_{2\theta }$ should be $U(2)$. The
generalization to other cases is straightforward.

This argument looks quite natural but has a critical loophole. The
phenomenon looks pathological but has its origin in the very nature of the
type II von Neumann algebra.

We first note that by combining $n$ and $m$ one may construct an arbitrarily
small number in the form $n+m\theta $. Let $\theta _{1}=n_{1}+m_{1}\theta $
and $\theta _{2}=n_{2}+m_{2}\theta $ be such small numbers and $P_{\theta
_{i}}$ be the corresponding PR projections of type (ii). We note that the PR
projection remains a projection operator when we parallel transport $f$ and $%
g$ in the $U$ direction. It can be achieved by replacing $U$ by 
$U\,e^{2\pi i \lambda}$ in (\ref{PR2}). 
We denote as $P_{\theta _{2}}^{\prime }$ the
projector which is translated from the origin more than 
$\lambda \geq|\theta _{1}+\theta _{2}|$ and keep the $%
\epsilon $ parameters sufficiently small compared to the $\theta $'s. 
Then one may easily confirm that 
$P_{\theta _{1}}P_{\theta _{2}}^{\prime }=0$.
This is because there is no overlap of the supports of the functions
$f$ and $g$ in these projectors even after the application of 
the operator $V^{m_i}$ which will cause translation of $\theta_i$.
This means that the projection operator for $\theta =\theta _{1}+\theta _{2}$
can be decomposed into mutually orthogonal but not necessarily the same type
of branes. Therefore, our argument that the branes with $k(p+q\theta )$ with
$p,q$ coprime must be split to $k$ identical $(p+q\theta )$ branes was too
naive.

The situation is actually much more intricate. By repeating the argument, we
have to conclude that for arbitrarily large $N$, one may divide the
projection operator in the form, $\mathcal{U}P_{\theta }\mathcal{U}%
^{-1}=\sum_{r=1}^{N}P_{\theta _{r}}$ for some unitary transformation $%
\mathcal{U}$ with $P_{\theta _{r}}$ mutually orthogonal. In other words,
although the object corresponding to a D0-brane has finite size as we
mentioned, it can be divided into an arbitrary number of tiny branes%
\footnote{%
Such behavior reminds us of the instability of the (super)membrane \cite
{dWLN}.}. This is related to the mathematical fact that there is no smallest
unit in the type II von Neumann factor.

Such a discussion seems to imply the instability of the system which is not
present in the fuzzy torus or the lattice calculation. There are some hints
which may remedy such disease from the physical side. One point is that we
should not forget about the existence of the regularization parameter $%
\epsilon $ in the PR projection. Mathematically, it may be arbitrarily small
as long as it is non-zero. However, from the physical viewpoint, it provides
the lower bound to make such solutions stable. Such smallest length
parameter may be identified with the lattice spacing. Another (maybe
related) point is that the linear mass spectrum in (\ref{spectrum2}) is an
approximation valid only for large $B_{eff}$ in (\ref{BImass}). When we
consider the tiny branes, such an approximation is not valid and we have to
come back to the original definition (\ref{BImass}) where they have finite
mass.

The occurance of such phenomena seems not to be restricted to the NC torus.
Indeed it came from the fact that the D0-branes occupy a finite size in the
compact space and the ratio between them is irrational. On this ground, we
may conjecture that the instability may occur universally in tachyon
condensation of D-branes which wrap any compact space. We hope to come back
to this problem in a future paper.

\bigskip \noindent \textbf{Note added:} After completing our calculation, we
noticed \cite{schnabl} on the net which mentioned the Powers-Rieffel
projector and discussed its physical consequences. \bigskip

\begin{center}
\noindent{\large \textbf{Acknowledgments}}
\end{center}
I.B. is supported in part by a DOE grant
DE-FG03-84ER40168. 
H.K. and T.T. are supported by JSPS Research Fellowships for Young
Scientists. 
Y.M. is supported in part by Grant-in-Aid (\# 09640352) and in
part by Grant-in-Aid (\# 707) from the Ministry of Education, Science,
Sports and Culture of Japan. 
Both Y.M. and I.B. thank the JSPS and the NSF for making
possible the collaboration between Tokyo University and USC through the
collaborative grants, JSPS(US-Japan coorperative science program), 
NSF-9724831. Y.M. is grateful to the Caltech-USC Center for hospitality.
We would like to thank A. Kato, D. Minic, E. Ogasa, 
S. Terashima, N. Warner and E. Witten for valuable discussions.


\end{document}